\documentclass[aps,prb,floats,twocolumn,floatfix,showpacs]{revtex4}
\usepackage{psfig,epsf,graphicx}
\usepackage{amssymb}
\usepackage{amsmath}
\usepackage{eepic}

\newlength{\piclen}
\begin{document}

\title{Nonequilibrium transport through parallel double quantum dots in the
  Kondo regime}

\author{Y.-F.\ Yang and K.\ Held}
\affiliation{Max-Planck-Institut f\"ur
Festk\"orperforschung, 70569 Stuttgart, Germany}

 \date{\today}

\begin{abstract}
We extend a perturbative, nonequilibrium renormalization group 
approach to multi-orbital systems and apply it for studying 
transport through two parallel quantum dots coupled electrostatically.
In general, the conductance shows
pronounced Kondoesque peaks at three voltages.
One of these peaks disappears if,
as in some experiments, one of the dots is 
decoupled from one of the two leads.
For an asymmetric coupling to the leads,
also negative differential conductances 
are possible.
This is a genuine nonequilibrium effect,
accompanying the Kondoesque peaks. Moreover,
a 
criterion to distinguish  spin and orbital Kondo effect in such a system is discussed.

\end{abstract}


\maketitle

\section{Introduction}
Kondo physics in quantum dots has been
under intense investigation in recent years.
As was predicted theoretically,\cite{KondoQdotTheory,Meir93a}
the Kondo effect can lift the Coulomb blockade at low temperatures,
leading to a  conductance peak at zero bias.\cite{KondoQdotExp}
In such experiments, often more than  one orbital is involved.
This is unavoidable, if the temperature or Kondo temperature is larger or 
comparable to the  mean orbital splitting.
Some nanostructures are also intendedly 
constructed in such a way, for example,
two parallel quantum dots  coupled electrostatically.\cite{Weisexp,Holleitner,Cdot}
The latter allows for a controlled study of the interplay of
spin and orbital (upper/lower dot) degrees of freedom.

On the theoretical side, the  $N$-fold  degenerate
situation where all  $N$ (spin and orbital) 
levels of the Anderson model have the same 
one-particle energy $\varepsilon$, Coulomb interaction $U$,
and hopping amplitude $t$  to the leads
is best understood.
At strong coupling,
the Schrieffer-Wolff transformation \cite{Schrieffer66a}
then leads to the Coqblin-Schrieffer model \cite{Coqblin69a}
in terms of the exchange coupling
$J$.
This model  has SU(N) symmetry for the rotation of orbitals and spins
and, with increasing degeneracy $N$, leads to a strongly enhanced  Kondo temperature
$T_K^{\rm SU(N)} \approx D e^{-1/ (N \rho_0 J)}$ ($\rho_0$: 
density of states in the leads; $D$:
bandwidth of the leads;\cite{footnote1}
we set $k_B=\hbar=e=1$ unless these constants appear explicitly in the equations).

Of course, we cannot expect a real quantum dot to
have SU(N) symmetry, unless special arrangements are made. \cite{Noteefforts}
Hence, quite an effort
was devoted  in the literature\cite{MlevelQdot,Izumida98a,deltalit,Borda03a,Eto04a} to study the situation 
where either a magnetic field or a difference in orbital energy $\delta$
splits $N=4$ levels into two doubly-degenerate subsets, 
hence breaking the SU(4) symmetry.
If $\delta\!\ll\! T_K^{\rm SU(4)}$ the low-energy physics 
 still resembles that of the SU(4)-symmetric model.
In contrast for $\delta\!\gg\! T_K^{\rm SU(4)}$ one of the levels 
drops out of the scaling procedure; the
low-energy physics is that of the
usual  SU(2)-symmetric Kondo model.
In between, there is a crossover  region.\cite{Borda03a}

It is also 
unavoidable that the 
coupling
constants $J$ are orbital-dependent in experiment.
Numerical renormalization group (NRG) calculations\cite{Izumida98a,Borda03a}
suggest that for small enough $\delta$
both orbitals are screened at the same energy,
in agreement with scaling analyses showing
a robust (marginal) SU(4) fix point.\cite{Eto04a,Ye,Borda03a}

\begin{figure}[b]
{\includegraphics[width=7.cm]{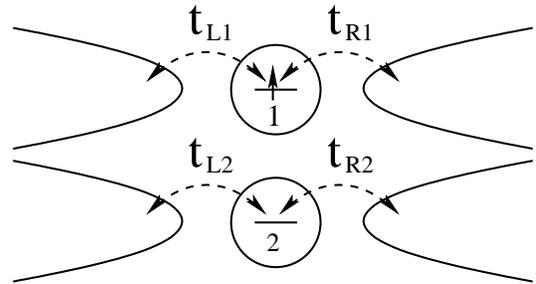}}
\caption{Sketch of the parallel double quantum dot configuration.
Each dot is coupled to its own leads; the coupling between the the
dots is only via the Coulomb interaction (not indicated in the Figure).
\label{FigTwodot}}
\end{figure}

Let us also distinguish here between orbital conserving and non-conserving
 dot-lead couplings.
In our paper, we consider the former, experimentally realized by separate leads for
each dot,\cite{Weisexp} see  Figure \ref{FigTwodot}.
The non-conserving case is obtained if the quantum dots 
couple to a single lead like 
in Ref.\ \onlinecite{Holleitner}.
Then 
the hopping processes do not conserve the orbital  quantum number,
which can give rise to SU(2) Kondo physics even if more than one degenerate level is involved,
see Ref.\ \onlinecite{Lopez} and references therein.

So far, we only discussed  equilibrium physics.
A finite voltage $V$ leads however to genuine nonequilibrium effects.
For example, if $V\!\approx\!\delta\!\gg\! T_K$ (the Kondo temperature) scattering
between the two subsets of levels becomes relevant again.
A Kondoesque resonance develops  at $V\!\approx\!\delta$,
visible e.g.\ as a peak in the conductance. But, at the same time, 
a genuine effect of nonequilibrium is the presence of
decoherence processes induced by the
finite current. This decoherence cuts off
the flow to the strong coupling fix point
even for temperatures $T\!\ll\! T_K$.\cite{Meir93a,Rosch01a,Paaske04a}
Then, physical quantities such as the differential conductance depend
strongly on the configuration of the system like the asymmetry of the
hopping $t$ between quantum dot and  leads. 

Nonequilibrium Kondo physics  is  not as thoroughly investigated 
theoretically because
standard approaches for the Anderson impurity model
like NRG are not applicable.
Rosch and coworkers\cite{Rosch03a} recently developed a perturbative, nonequilibrium renormalization group (RG) method
 to address nonequilibrium Kondoesque physics.\cite{Kehrein04}
This approach is similar to Anderson's\cite{PoorMan}  poor man's scaling approach but 
includes nonequilibrium effects caused by the finite
current through the system. In nonequilibrium, the low energy physics
cannot be absorbed properly by a pure renormalization of the bare
couplings. It is necessary to include the frequency dependence of the 
 vertex functions.

The outline of the paper is as following:
In Section \ref{method},
we extend the perturbative RG of Rosch {\em et al.} \cite{Rosch03a} to many orbitals,
supplemented by the calculation of the decoherence rates  in the  Appendix.
In Section \ref{Sec2Level}, we 
apply this method to the parallel double quantum dot system,
with each dot being coupled to separate leads.
Thereby,  the first part, Section  \ref{SecSimp},
 discusses briefly
the simplification of the  RG Eqs.\ for this  special  application.
The following Sections present our  results:
Section \ref{SecVertex} shows
the renormalized vertex function, primarily of theoretical interest,
Section \ref{SecGamma}  the increase of the decoherence
rates with voltage, and Section  \ref{SecOccupation} the change
of the occupation in non-equilibrium.
The reader who is mainly interested in the differential conductance
may also directly turn to Sections \ref{SecSym} and  \ref{SecAsym}
which present our results for this physical observable in the case of
symmetric and asymmetric coupling to the
 two leads, respectively.
In the case of symmetric couplings (Section \ref{SecSym})
we generally obtain two orbital Kondoesque peaks at 
$V\approx\pm\delta$ in the differential conductance of both dots 
and an additional spin Kondo
peak at $V=0$ for the dot which is lower in energy.
Depending on the parameters, 
the peak at $V=0$  can be strongly suppressed and hardly discernible.
The main findings of 
Section \ref{SecAsym} for asymmetric couplings are:
(i) in very asymmetric situations the
Kondoesque peaks at $V\approx-\delta$ can disappear
and  (ii) negative differential conductances
are possible for $V\gtrsim\delta$.
In addition to these non-equilibrium results,
 the linear (equilibrium) conductance
is analyzed in
Section \ref{SecLinear}.
A  summary of the results can be found in
Section \ref{SecConclusion}.

\section{Nonequilibrium perturbative RG}
\label{method}
Starting point  for  modeling
a quantum dot with $N$-levels should be the
 Anderson impurity model:
\begin{eqnarray}
H_{\rm AIM}&=&H_0+\sum_{\lambda
mk}(t^{\phantom{+}}_{\lambda m}c^+_{mk\lambda}d_{m\phantom{k}\!}^{\phantom{+}}+{\rm H.c.})\nonumber \\
&&+\sum_m\varepsilon^{\phantom{+}}_m d^+_m d^{\phantom{+}}_m+\sum_{m\neq m'}U_{mm'}n_m
n_{m'}.
\label{HAIM}
\end{eqnarray}
Here $d^+_m$ and $d_m^{\phantom{+}}$ ($c^+_{mk\lambda}$ and $c_{mk\lambda}$) are
 creation and annihilation operators 
for electrons in the dot (lead $\lambda$);
$n_m=d^+_m d_m^{\phantom{+}}$; $\varepsilon_m$ describes the one-particle energy 
of level $m$ with $m=\{\sigma,i\}$ subsuming the spin and orbital
index; $U_{mm'}$ is the Coulomb repulsion within the
dot. The  levels of the dot hybridize via $t_{\lambda m}$ with
non-interacting leads $\lambda$ described by (later, we simplify $\epsilon_{\lambda m k}=\epsilon_k$)
\begin{equation}
H_0=\sum_{\lambda m k} \epsilon^{\phantom{+}}_{\lambda m k}c^+_{\lambda m k}
c^{\phantom{+}}_{\lambda m k}. 
\end{equation}

In this paper, we assume each quantum dot level $m$ to couple to its own lead channel which is
for example the case if the different levels correspond to
different quantum dots like in Ref.\ \onlinecite{Weisexp} 
(see   also Figure  \ref{FigTwodot}
of the present paper as an illustration). But
in other situations it is possible
to move an electron phase coherently from one level $m$ via
the leads to another level $m^\prime$. This goes beyond
Hamiltonian (\ref{HAIM}).

While the Anderson impurity model allows us to make
contact with experiment (estimating parameters),
we address in the following  the parameter regime 
$\epsilon_m\ll \mu_{\lambda}\ll U_{mm'}+\epsilon_m$
for all $m$,$m'$,$\lambda$ ($\mu_{\lambda}$: chemical potential in lead $\lambda$)
so that charge fluctuations are suppressed. 
Then, with  $\sum_{m}\langle n_m\rangle=1$ electron within the dot,
we can map Hamiltonian (\ref{HAIM}) onto the subspace with one electron in the dot
by a Schrieffer-Wolff transformation,\cite{Schrieffer66a} i.e.,
onto the general effective Kondo Hamiltonian (neglecting a potential
scattering term)
\begin{widetext}

\begin{eqnarray}
H&= & H_0+ \sum_{m} \epsilon_m X^{mm} 
+\!\!\!\! \sum_{\lambda_1m_1k_1;\lambda_2 m_2 k_2}(J^{\lambda_1\lambda_2}_{m_1m_2}X^{m_1m_2}c^+_{\lambda_2
  m_2k_2}c_{\lambda_1 m_1k_1}^{\phantom{+}}+\tilde{J}^{\lambda_1\lambda_2}_{m_1m_2}X^{m_1m_1}c^+_{\lambda_2
  m_2k_2}c_{\lambda_1 m_2k_1}^{\phantom{+}}).
\label{Kondo}
\end{eqnarray}

\end{widetext}
Here, the Hubbard operators\cite{HubbardX} $X^{m_1m_2}$ represent the scattering of the local state from level $m_2$
to level $m_1$. When the lead's band edge $D>|\varepsilon_m|$,
the local level  $\varepsilon_m$ of the Anderson impurity model is renormalized to 
 $\epsilon_m$ in the Kondo model;\cite{Hewson}
$J^{\lambda_1\lambda_2}_{m_1m_2}=t^*_{\lambda_1m_1}t^{\phantom{*}}_{\lambda_2m_2}(\frac{1}{U_{m_1m_2}
+\epsilon_d}-\frac{1}{\epsilon_d})$ and
$\tilde{J}^{\lambda_1\lambda_2}_{m_1m_2}=(1-2\delta_{m_1m_2})t^*_{\lambda_1m_2}t^{\phantom{*}}_{\lambda_2m_2}\frac{1}{U_{m_1m_2}
+\epsilon_d}$
where  $\epsilon_d$ is the average level energy,
neglecting the level splitting which is much smaller than $U,\epsilon_d$.
Note, that for $m_1=m_2$ both $J$ and $\tilde{J}$ yield
the same kind of contribution. The advantage of having this
term twice (i.e.,
distributed to $\tilde{J}^{\lambda_1\lambda_2}_{m_1m_1}$ and ${J}^{\lambda_1\lambda_2}_{m_1m_1}$) is the simplification of the following equations.
The same distribution of the $m_1=m_2$ contribution
is usually also employed for the Coqblin-Schrieffer model
 which is the SU(N) symmetric version
of Hamiltonian (\ref{Kondo}) [$J$ and $\tilde{J}=J/N$
are then level-independent].\cite{Coqblin69a}

\begin{figure}[tb]
{\includegraphics[width=8.6 cm]{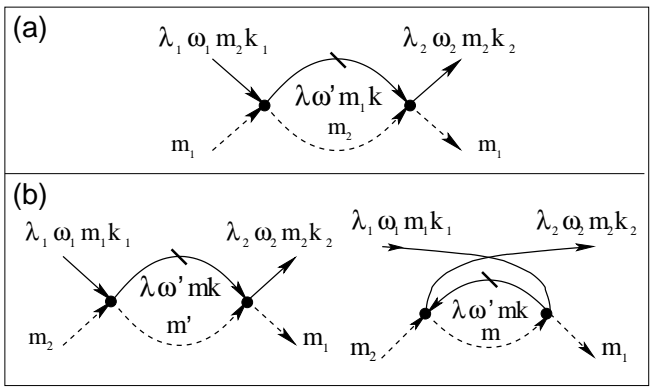}}
\caption{
One-loop Feynman diagrams for the renormalization
of the vertex functions (a) $\tilde{g}_{m_1m_2}^{\lambda_1\lambda_2}(\omega)$
 and (b)  ${g}_{m_1m_2}^{\lambda_1\lambda_2}(\omega)$.
The full line denotes the electron and the dashed the localized
state of the Kondo model.
Note that the  second (hole) diagram of (b) does not contribute for
  (a) since it cancels with the
corresponding first (particle) diagram.
For the vertex renormalization, the electron line  is restricted
 to a small interval $|{\rm d}D|$ at 
the band edges,
as indicated by the dash.
\label{figFD}
}
\end{figure}

Let us now generalize the perturbative RG
approach of Ref.\ \onlinecite{Rosch03a} to the multi-orbital
Hamiltonian (\ref{Kondo}). The base quantity is the
vertex function which -at finite voltages or currents- becomes
frequency dependent. Like in the poor man's scaling approach,
we integrate out high energy degrees of freedom 
in the intervals $[-D,-D+|{\rm d}D|]$ and $[D-|{\rm d}D|,D]$, resulting in 
 a renormalization
of the vertex. The 
Feynman diagrams for this renormalization are shown in Figure\ \ref{figFD} to
 lowest order (one-loop RG). To directly apply standard perturbation theory,
the localized states of the Anderson impurity model 
can be expressed by pseudo Fermions, see Ref.\ \onlinecite{Rosch03a}.
Starting point of the RG scheme is the unrenormalized ${k}$-independent vertex 
${g}_{m_1m_2}^{\lambda_1\lambda_2}(\omega)=\rho_0 {J}_{m_1m_2}^{\lambda_1\lambda_2}$, 
$\tilde{g}_{m_1m_2}^{\lambda_1\lambda_2}(\omega)=\rho_0 \tilde{J}_{m_1m_2}^{\lambda_1\lambda_2}$, where
we neglect the energy dependence of $\rho_0$. Here, 
the frequency of the localized state is
approximated by $\epsilon_m$
so that the vertex depends only on one
$\omega$ (the average of
incoming and outgoing electron frequency).
This is only possible in the weak coupling regime.

\begin{figure}[tb]
\vspace{.3cm}

\includegraphics[width=8.6 cm]{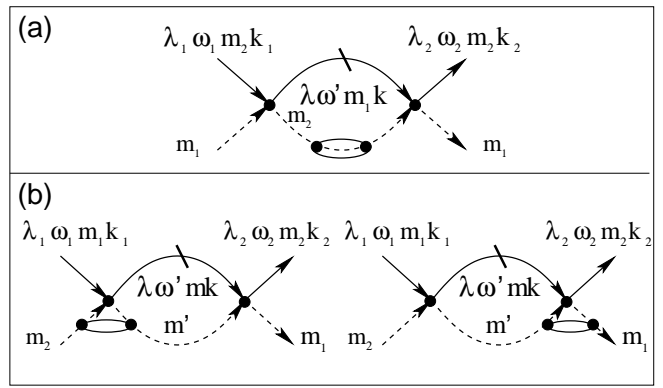}
\caption{
Vertex corrections and self energy diagrams 
beyond the one-loop approximation
of Figure \ref{figFD}. 
\label{FigHighOrder}
}
\vspace{.2cm}
\end{figure}

For the nonequilibrium situation,
the calculation of the diagrams of Figure\ \ref{figFD} requires 
in principle the usage of  Keldysh \cite{Keldysh} Green functions.
Rosch {\em et al.}\cite{Rosch03a} argued however
that, to  leading order in $1/\ln(V/T_K)$, it is sufficient to
keep track of the real part of ${g}_{m_1m_2}^{\lambda_1\lambda_2}(\omega)$,
as  indicated by perturbation theory.\cite{Paaske03a} This
gives rise to the same kind of vertex
renormalization  per energy interval $\ln D$ as in the
poor man's scaling approach, i.e, at 
$T=0$:
\begin{equation}
\frac{1}{2}\frac{\partial}{\partial\ln
D}\int^{D}_{-D}d\epsilon\frac{{\rm sign}(\epsilon)}{\epsilon-\Delta\omega}. \label{pole1}
\end{equation}
Here, $\Delta \omega$ is given by the chemical potential 
and the change of energy of the localized state in Figure\ \ref{figFD}.
 A finite
temperature, smears out the sharp step  of the integrand in Eq.\ (\ref{pole1}), cutting off the emerging logarithm.
Another cut-off is induced by the finite voltage, driving
the system out of equilibrium. This nonequilibrium effect
originates from higher order diagrams such 
as those of Figure\ \ref{FigHighOrder}.
These diagrams
give rise to an imaginary part of the self energy 
and vertex corrections, even at low $\omega$'s for the 
nonequilibrium situation. Generally, this will result in the following 
modification of the
pole in Eq.~(\ref{pole1}):
\begin{eqnarray}
\lefteqn{\frac{1}{2}\frac{\partial}{\partial\ln
D}\int^{D}_{-D}d\epsilon\frac{{\rm sign}(\epsilon)}{\epsilon-\Delta\omega+i
  \gamma {\rm sign}(\epsilon)}}\nonumber\\\;\;\;\;&&\;\;\;\;\approx \Theta(D-\sqrt{\Delta\omega^2+\gamma^2}) \equiv \Theta^{\gamma}_{\Delta\omega},
\label{pole2}
\end{eqnarray}
where $\Theta(x)=1(0)$ for $x\ge 0(<0)$.

\begin{figure}[tb]
\includegraphics[width=8.6 cm,height=2.2cm]{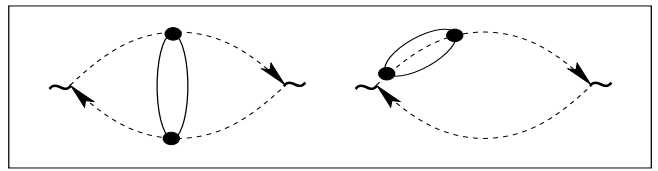}
\caption{
A similar combination of 
vertex corrections and self energy diagrams 
as in Figure \ref{FigHighOrder} emerges if we
calculate generalized spin/orbit susceptibilities. 
\label{FigSus}
}
\end{figure}

For the proper calculation of the low energy cut-off
$\gamma$, we would need to calculate diagrams 
as in Figure\ \ref{FigHighOrder}. This goes beyond the scope of
the present paper. As in Ref.\  \onlinecite{Rosch03a},
we follow a heuristic approach instead.
To this end, first note that similar combinations
of vertex corrections and self energy inclusions
occur when calculating generalized susceptibilities
for level $m_1$ and $m_2$;
see Figure\ \ref{FigSus}.
Secondly,  an incorrect
prefactor of $\gamma$
is not affecting the results
to leading order in $1/{\rm ln} (V/T_K)$.
Hence, we can 
crudely estimate $\gamma$
by  the decoherence rates
$\gamma_{m_1m_2}$ of the corresponding susceptibilities. For example, we can 
take the maximal  $\gamma_{m_1m_2}$ of the levels involved
(or temperature if higher):
 $\gamma\equiv \gamma_{m_1m_2m_3}=
\max\{T,\gamma_{m_1m_2},\gamma_{m_2m_3}\}$
where $m_1$, $m_2$ and $m_3$ denote the
incoming, intermediate, and final local state.
In our calculation for two levels in Section
\ref{Sec2Level} with SU(2) spin symmetry,
only two 
of the orbitals $i$ ($m=\{\sigma,i\}$) can be different,
and the vertices are symmetric w.r.t.\ the spin indices.
We then actually
use the intra-orbital rate $\gamma=\max\{T,\gamma_{ii}\equiv\gamma_i\}$
as a cut-off for diagrams where only one orbital
is involved and   the inter-orbital
$\gamma=\max\{T,\gamma_{12}\}$ whenever
both orbitals are involved.
The calculation of $\gamma_{m_1m_2}$
via the  susceptibilities is presented in the Appendix.

Having identified the cut-off $\gamma_{m_1m_2m_3}$, we can now
write down the RG equation for
the vertex functions $\tilde{g}_{m_1m_2}^{\lambda_1\lambda_2}(\omega)$
and ${g}_{m_1m_2}^{\lambda_1\lambda_2}(\omega)$, corresponding
to Figure \ref{figFD}:

\begin{widetext}
\vspace{-.15cm}
\begin{eqnarray}
\frac{\partial{\tilde{g}_{m_1m_2}^{\lambda_1\lambda_2}(\omega)}}{\partial{\ln{D}}}
\!\!&=&\sum\limits_{\lambda}
g_{m_2m_1}^{\lambda_1\lambda}(\omega+\frac{\epsilon_{m_1}-\epsilon_{m_2}}{2})g_{m_1m_2}^{\lambda\lambda_2}(\omega+\frac{\epsilon_{m_1}-\epsilon_{m_2}}{2})\Theta^{\gamma_{m_2m_1m_2}}_{\omega+\epsilon_{m_1}-\epsilon_{m_2}-\mu_{\lambda}}, \label{scaling1}\nonumber
\\
\frac{\partial{g_{m_1m_2}^{\lambda_1\lambda_2}(\omega)}}{\partial{\ln{D}}}
\!\!&=&\!\!-\sum\limits_{m\lambda} g_{m_1m}^{\lambda_1\lambda}(\omega-\frac{\epsilon_{m_2}-\epsilon_m}{2})
g_{mm_2}^{\lambda\lambda_2}(\omega-\frac{\epsilon_{m_1}-\epsilon_m}{2}) 
\Theta^{\gamma_{m_2mm_1}}_{\omega-(\epsilon_{m_1}+\epsilon_{m_2}-2\epsilon_{m})/2-\mu_{\lambda}}
\nonumber \\ 
&&\!\!+\sum\limits_{\lambda}
g_{m_1m_2}^{\lambda_1\lambda}(\omega) \big[\tilde{g}_{m_1m_2}^{\lambda\lambda_2}(\omega\!-\!\frac{\epsilon_{m_1}\!-\!\epsilon_{m_2}}{2})\Theta\!^{\gamma_{m_2m_1m_1}}_{\omega-(\epsilon_{m_1}\!-\epsilon_{m_2})\!/\!2\!-\!\mu_{\lambda}}\!\!-\tilde{g}_{m_2m_2}^{\lambda\lambda_2}(\omega\!-\!\frac{\epsilon_{m_1}\!-\!\epsilon_{m_2}}{2})\Theta\!^{\gamma_{m_2m_2m_1}}_{\omega-(\epsilon_{m_1}\!-\epsilon_{m_2}\!)\!/\!2\!-\!\mu_{\lambda}}\!\big]
\nonumber \\
&&\!\!+\sum\limits_{\lambda}
g_{m_1m_2}^{\lambda\lambda_2}(\omega) \big[\tilde{g}_{m_2m_1}^{\lambda_1\lambda}(\omega\!+\!\frac{\epsilon_{m_1}\!-\!\epsilon_{m_2}}{2})\Theta^{\gamma_{m_2m_2m_1}}_{\omega+(\epsilon_{m_1}\!-\epsilon_{m_2}\!)\!/\!2\!-\!\mu_{\lambda}}\!\!-\tilde{g}_{m_1m_1}^{\lambda_1\lambda}(\omega\!+\!\frac{\epsilon_{m_1}\!-\!\epsilon_{m_2}}{2})\Theta\!^{\gamma_{m_2m_1m_1}}_{\omega+\!(\epsilon_{m_1}\!-\epsilon_{m_2}\!)\!/\!2\!-\!\mu_{\lambda}}\!\big]. \label{scaling2}
\end{eqnarray}
\vspace{-.15cm}
\end{widetext}
The equation for $\tilde g$ can be obtained from
Figure \ref{figFD}a. 
The first line of the equation for $g$
corresponds to the right diagram of Figure \ref{figFD}b;
the second and third line  stem from the 
particle and the hole diagram of Figure \ref{figFD}b
with  $m=m_2$ and  $m=m_1$, respectively.
Note that the different frequencies of Figure \ref{figFD}
are related by energy conservation
and that  $\omega=(\omega_1+\omega_2)/2$ 
\cite{NoteDiff}.
The
$k$-integration (note that $k$ is not conserved)
over the band edge states 
results in a factor $\rho_0$
which is already included in $g$.
The electron states in the leads range to $\pm D$
relative the the respective chemical potential $\mu_{\lambda}$.

The renormalized vertex functions of Eq.\ (\ref{scaling2})  have to be calculated self-consistently  
together with $\gamma_{m m'}$ [Eq.\ (\ref{EqGamma}) of the Appendix]
 which determine the cut-offs $\gamma_{m_1 m_2 m_3}$.

\subsection*{Conductance}
\label{conductance}

To study the most important physical quantity, the conductance,
 we employ the scattering T-matrix
$\hat{T}_{m_1m_2}^{\lambda_1\lambda_2}(k_1,k_2)$  for the process
moving  an electron from
state $\{k_1m_1 \}$ in lead $\lambda_1$ to state $\{k_2m_2 \}$ in lead
$\lambda_2$. This process 
necessarily also involves the
local state which changes in the opposite
direction. The change of the local state can be expressed
via spin and orbital operators in the T-matrix
(see e.g.\ Ref.\ \onlinecite{Hewson}).
By $\langle|\hat{T}_{m_1m_2}^{\lambda_1\lambda_2}(k_1,k_2)|^2\rangle$, we then
denote the thermal average of the T-matrix
w.r.t.\ the local state configurations. The steady state condition requires
in terms of this averaged T-matrix:
\begin{eqnarray}
\!\!\!\!\!\!\!\!\!\!\lefteqn{\!\!\!\!\!\!\!\sum\limits_{\lambda_1\lambda_2;k_1k_2}\!\!\!\!\!\!\!
\langle |\hat{T}_{m_1m_2}^{\lambda_1\lambda_2}(k_1,k_2)|^2\rangle
f_{\lambda_1}\!(\epsilon_{k_1}\!)[1\!-\!f_{\lambda_2}\!(\epsilon_{k_2})]\!=}\nonumber \\&& \sum\limits_{\lambda_1\lambda_2;k_1k_2}\!\!\!\!\!\!\! \langle|\hat{T}_{m_2m_1}^{\lambda_2\lambda_1}(k_2,k_1)|^2\rangle
f_{\lambda_2}\!(\epsilon_{k_2}\!)[1\!-\!f_{\lambda_1}\!(\epsilon_{k_1})],\label{Tmatrix}
\end{eqnarray}
where  $f_{\lambda}(\epsilon)=1/(e^{\beta (\epsilon-\mu_{\lambda})}+1)$ is the Fermi 
function for lead $\lambda$. 
A difference to Ref.\ \onlinecite{Rosch03a,Rosch05a} is that
we do not take into account a $\gamma$-broadening
of the spectral function. This   is effectively
described  by a broadening of the Fermi functions in
 Ref.\ \onlinecite{Rosch03a,Rosch05a}, resulting in
corrections to  subleading order in $1/\ln(V/T_K)$.

Since we consider the situation where 
either  the voltage-induced decoherence rate
or the
temperature is
larger than the Kondo temperature
the renormalization process  is cut off, instead of flowing to strong coupling.
Therefore, we are at weak coupling and we 
can calculate the scattering matrix in lowest order:
 $\hat{T}^{\lambda_1\lambda_2}_{m_1m_2}(k_1,k_2)=
\langle \lambda_2m_2k_2 | H_{\rm int} |{\lambda_1 m_1 k_1}\rangle$, where
 $H_{\rm int}$ denotes the interaction between the lead electrons and the local states
via $J$ and $\tilde J$, i.e., the last two terms of Eq.\ (\ref{Kondo}).
Averaged over the local states, 
this yields:
\begin{eqnarray}
\lefteqn{\!\!\!\!\!\!\!\!\langle|\hat{T}^{\lambda_1\lambda_2}_{m_1m_2}(k_1,k_2)|^2\rangle = 
(1-\delta_{m_1m_2})\langle n_{m_2}\rangle 
 g^{\lambda_1\lambda_2}_{m_1m_2}(\omega)^2/\rho_0^2  \label{Tmatrix1}} \nonumber \\
&&\;\;\;\;+ \delta_{m_1m_2}\sum\limits_{m\neq m_1} \langle
 n_{m}\rangle\tilde{g}^{\lambda_1\lambda_2}_{m_1m}(\omega)^2/\rho_0^2 \nonumber \\
&&\;\;\;\;+ \delta_{m_1m_2}\langle n_{m_1}\rangle
 (g^{\lambda_1\lambda_2}_{m_1m_2}(\omega)+\tilde{g}^{\lambda_1\lambda_2}_{m_1m_2}(\omega))^2/\rho_0^2.
\label{Tmatrix2}
\end{eqnarray}
 Here, we have replaced the bare coupling
by the renormalized one, i.e., $J\rightarrow g(\omega)/\rho_0$
with
$\omega\equiv(\epsilon_k+\epsilon_{k'})/2$;  $\langle n_m \rangle$ is the average occupation
of local level  $m$.

In a first step  $g$ and $\tilde g$
are calculated by Eq.\  (\ref{scaling2}),
which is solved self-consistently together with the
cut-off $\gamma_{m_1 m_2}$  [Eq.\ (\ref{EqGamma})].
With these  $g$'s and $\tilde g$'s,
 the occupations $\langle n_m\rangle$ can be determined
via  Eqs.\
(\ref{Tmatrix}) and (\ref{Tmatrix1})
and the constraint 
$\sum_{m}\langle n_m \rangle =1$.

From the  T-matrix  [Eq.\ (\ref{Tmatrix1})],
 we can then calculate  the  current 
from lead  $\lambda$ and orbital $m$  as the
difference between scattering out of and into the lead:
\begin{eqnarray}
\!\!I_{\lambda m}\!\! &=-2\pi \!\!\!\!\!\!\!\!\sum\limits_{\stackrel{\scriptstyle 
\lambda'm'(\neq\lambda m)}{k k'}}\!\!\!\!\!\!\!  &\delta(\epsilon_k\!-\!\epsilon_{k'}+\epsilon_{m'}\!-\!\epsilon_m) \nonumber \\ 
  \!\!  \!& &\!\!\!\!\!\!\!\big\{f_{\lambda}\!(\epsilon_{k}\!)[1\!-\!f_{\lambda'}(\epsilon_{k'})]
\langle|\hat{T}_{mm'}^{\lambda\lambda'}(k,k')|^2\rangle\nonumber \\ 
&&\!\!\!\!\!\!\! -f_{\lambda'}\!(\epsilon_{k'}\!)[1\!-\!f_{\lambda}(\epsilon_{k})]
\langle|\hat{T}_{m'm}^{\lambda'\lambda}(k',k)|^2\rangle\big\}.\label{current}
\end{eqnarray}

The differential conductance follows as
$G_{\lambda m}=\partial
I_{\lambda m}/\partial V$.

\section{Two  parallel quantum dots}
\label{Sec2Level}

Let us now apply the perturbative RG equations to 
the situation of two  parallel quantum dots each coupled to two leads,
see Figure \ref{FigTwodot}  of the introduction.
For the Anderson impurity model, this dot-lead coupling is described
by  hopping
processes $t_{\lambda i}$; additionally
there are  intra-dot and inter-dot Coulomb repulsions
$U_{ii'}$.
We study here the corresponding Kondo model
which is restricted to the subspace with one electron in the dot.
Then, $g_{ii'}^{\lambda\lambda'}$
describes the combination of two hopping processes,
leaving the number of electrons in the dot unchanged.

\subsection{Simplification of the RG Eq.}
\label{SecSimp}

The two  parallel quantum dot case corresponds to quantum numbers
$m=\{i,\sigma\}$ with $\sigma=\uparrow(+),\downarrow(-)$ denoting the spin
and $i=1(+),2(-)$ the two dots (orbitals) with level energy
$\epsilon_{\{i\sigma\}}=-i\delta/2$ ($\bar{i}=-i$ in the following).

Due to the SU(2) spin symmetry in each dot,
we can reduce the number of  vertices we have to deal with.
Furthermore, some of the vertices are connected via scaling invariants.
Altogether this allows for the following reduction of vertices:
 $g_{i\sigma;i\sigma'}^{\lambda \lambda'}(\omega)\equiv g_i^{\lambda \lambda'}(\omega)/2$,
$g_{i\sigma;\bar{i}\sigma'}^{\lambda \lambda'}(\omega)\equiv g_{12}^{\lambda \lambda'}(\omega)/2$,
$\tilde{g}_{i\sigma,\bar{i}\sigma'}^{\lambda \lambda'}(\omega)\equiv-\tilde{g}^{\lambda \lambda'}_{\bar{i}}(\omega)/4$, and  -from the scaling
invariants- $\tilde{g}_{i\sigma;i\sigma'}^{\lambda \lambda'}(\omega)=(\tilde{g}_{{i}}^{\lambda \lambda'}(\omega)-g_i^{\lambda \lambda'}(\omega))/4$. 
The meaning of the redefined vertices becomes
clear if we write the Hamiltonian in terms of
the coupling constants
$J_{i(12)}$ and $\tilde{J}$ corresponding to
$g_{i(12)}$ and $\tilde g$:
\begin{eqnarray}
H&=&\sum\limits_{\lambda i\sigma k}(\epsilon_k-\mu_{\lambda})
c^+_{\lambda i\sigma k}c^{\phantom{+}}_{\lambda i\sigma
  k} -\frac\delta2(n_1-n_2)\\ \nonumber
 & & \!+\sum\limits_{\lambda\lambda^{\prime}i\sigma\sigma^{\prime}kk^{\prime}}\frac{J^{\lambda^{\prime}\lambda}_i}{2}\mathbf{S}_i\cdot
c^+_{\lambda i\sigma k}\mbox{\boldmath ${\tau}$}_{\sigma\sigma^{\prime}}
c_{\lambda^{\prime} i\sigma^{\prime}k^{\prime}}
\\ \nonumber
 & & \!+\sum\limits_{\lambda\lambda^{\prime}\sigma\sigma^{\prime}kk^{\prime}}\frac{J^{\lambda^{\prime}\lambda}_{12}}{2}(X^{\{1\sigma^{\prime}\}\{2\sigma\}}c^+_{\lambda 2\sigma k}c_{\lambda^{\prime}1\sigma^{\prime}k^{\prime}}\!+\rm H.c.)\\
& &\!+(n_1\!-\!n_2)\!\sum\limits_{\lambda\lambda^{\prime}\sigma kk^{\prime}}\!\frac{\tilde{J}^{\lambda^{\prime}\lambda}}{4}(c^+_{\lambda 1\sigma k}c_{\lambda^{\prime}1\sigma k^{\prime}}\!-\!c^+_{\lambda 2\sigma k}c_{\lambda^{\prime}2\sigma k^{\prime}}).\nonumber
\end{eqnarray}
Here, $\mathbf{S}_i$ denotes  the spin operators in orbital $i$,
$\mbox{\boldmath ${\tau}$}$ the vector of Pauli matrices,
 and $n_i=\sum_\sigma n_{i\sigma}$. Note that with only one electron in the dots, there is no need
to consider Hund's coupling.

Using the SU(2) symmetries in both orbitals and the
above scaling invariants, the scaling Eq.\ (\ref{scaling1}) 
become particularly simple for
a symmetric coupling to the leads (then,
$g_i(\omega)$, $\tilde{g}_i(\omega)$, and $g_{12}(\omega)$ do
not depend on the lead indices; we can hence drop the
$\lambda$, $\lambda'$ indices):
\begin{equation}
\begin{split}
\frac{\partial g_i(\omega)}{\partial\ln D}\!&=\!-\!\!\!\!\sum\limits_{\lambda=\pm 1}\![g^2_i(\omega)%
\Theta^{\gamma_i}_{\omega+\lambda
V/2}\!+\!g^2_{12}(\omega\!+\!{\scriptstyle \frac{\scriptstyle i \delta}{\scriptstyle 2}})\Theta^{\gamma_{12}}_{\omega+i(\delta+\!\lambda V/2)}], \\
\frac{\partial g_{12}\!(\omega)}{\partial\ln D}\!&=-g_{12}\!(\omega)\!\!\!\!\sum\limits_{i;\lambda=\pm 1}\!\!\! [%
{\scriptstyle{\frac{3}{4}}} g_i(\omega\!-\!{\scriptstyle \frac{i\delta}{2}})\!+\!{\scriptstyle \frac{1}{2}} \tilde{g}_i(\omega\!-\!{\scriptstyle \frac{i\delta}{2}})]\Theta^{\gamma_{12}}_{%
\omega+\frac{\lambda V\!-i\delta}{2}}, \\
\frac{\partial \tilde{g}_i(\omega)}{\partial\ln D}\!&=-\sum\limits_{\lambda =\pm 1}g^2_{12}(%
\omega+i\delta/2)\Theta^{\gamma_{12}}_{\omega+i(\delta+\lambda V/2)}.
\end{split}\label{scaling2level}
\end{equation}

Eq.\ (\ref{scaling2level}) has to be solved self-consistently
together with the decoherence rates of the Appendix which simplify for
the parallel double dot system to:
\begin{eqnarray}
\gamma_{i}&\!\!=\pi\!\sum\limits_{\lambda\lambda^{\prime}} \!\!&\!\int\! \!
d\omega\,\big\{g_i^2(\omega)f_\lambda( \omega)[1\!-\!f_{\lambda^{\prime}}(\omega)]
 \nonumber\\&&+g^2_{12}(\omega)f_{\lambda}(\omega\!+\!i
\frac{\delta}{2})[1\!-\!f_{\lambda^{\prime}}(\omega\!-\!i\frac{\delta}{2})]\big\}\\
\gamma_{12}&\!=\!\frac{\pi}{2}\!\sum\limits_{\lambda\lambda^{\prime}i} \!\!&\int\! \!d\omega\,\big\{[
\frac34g^2_i(\omega)\!+\!\tilde{g}^2_i(\omega)\}f_\lambda(\omega)[1-\!f_{\lambda^{
\prime}}(\omega)]\nonumber\\&&+g^2_{12}(\omega)f_{\lambda}(\omega\!+\!i\frac{\delta}{2})
[1\!-\!f_{\lambda^{\prime}}(\omega-i\frac{\delta}{2})]\big\}
\end{eqnarray}
These $\gamma_i$, or $T$ for $T>\gamma_i$,  cut off the $\Theta$-functions in Eq.\ (\ref{scaling2level}).

The equations for the
 occupations [Eq.\ (\ref{Tmatrix}), taking $m_1=m_2$ in Eq.\ (\ref{Tmatrix2})] and the current [Eq.\ (\ref{current})]
reduce to
\begin{equation}
\langle n_i \rangle \sum\limits_{\lambda,\lambda^{\prime}}\!\int\! d\epsilon
g^2_{12}(\epsilon\!-\!i\frac{\scriptstyle \delta}{\scriptstyle 2})f_\lambda(\epsilon)[1\!-\!f_{\lambda^{\prime}}(\epsilon\!-\!i \delta)]=(i\rightarrow \bar{i})
\end{equation}
and
\begin{eqnarray}
\!\!\!\!\!\!\!I_{i\sigma}\!\!&=\frac{\pi}{8}\!\int\!&\!\!\! d\omega \big\{[3 \langle n_i\rangle g^2_i(\omega)\!+\!\tilde{g%
}^2_i(\omega)][f_L(\omega)\!-\!f_R(\omega)] \label{current2} \\
&&\!\!+4\langle n_{\bar{\imath}}\rangle g^2_{12}(\omega\!+\!i\delta/2)f_L(\omega)[1\!-\!f_R(\omega\!+\!i%
\delta)] \nonumber \\
&&\!\!-4\langle n_{i}\rangle g^2_{12}(\omega\!-\!i\delta/2)f_R({\omega})[1\!-\!f_L(\omega\!-\!i\delta)]
 \nonumber \\ \nonumber
&&\!\!+4\langle n_{\bar{\imath}}\rangle g^2_{12}(\omega\!+\!i\delta/2)f_L(\omega)[1\!-\!f_L(\omega\!+\!i%
\delta)] \nonumber \\
&&\!\!-4\langle n_{i}\rangle g^2_{12}(\omega\!-\!i\delta/2)f_L(\omega)[1\!-\!f_L({\omega\!-\!i\delta})] \big\},\nonumber
\end{eqnarray}
respectively. In Eq.\ (\ref{current2}),
the first line stems from the current from left channel $i$ to right channel $i$. The 2nd and 3rd (4th and 5th) line
correspond to the current to right (left) channel $\bar{i}$.
The latter are balanced by similar terms
from the right lead so that the net current
from one orbital to the other is zero.

\subsection{Results: Renormalized vertex function}
\label{SecVertex}

\begin{figure}[tb]
{\includegraphics[width=8.6cm]{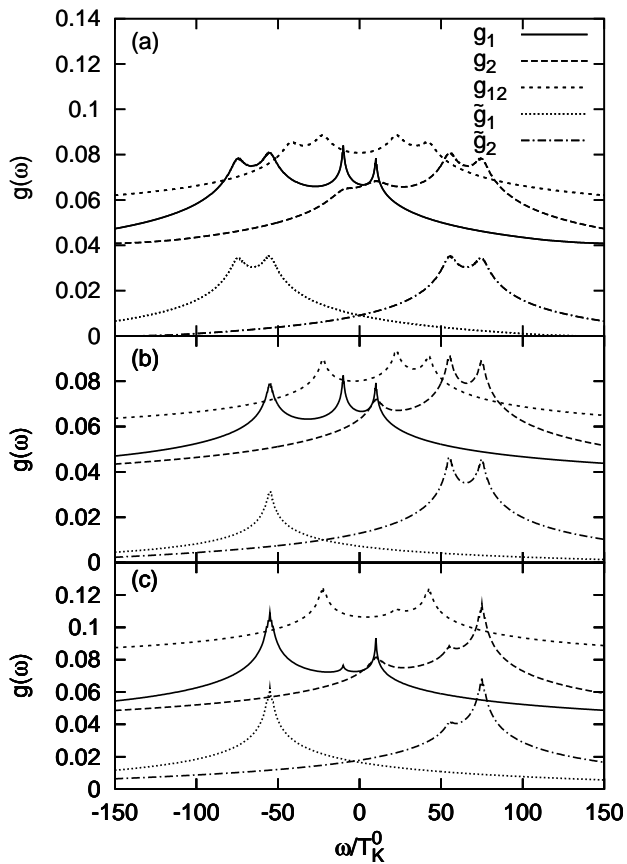}}
\caption{Renormalized vertex functions 
to the left lead
at $T/T^0_K=0.5,
  V/T^0_K=20,\delta/T^0_K=65$
($g_{i(12)}$ in the legend box stands for
$g_{i(12)}^{LL}$).
Part a) is for symmetric initial couplings $g_i^{\lambda \lambda'}=0.04$,
$g^{\lambda \lambda'}_{12} = 0.06$ for all $\lambda,\lambda'$.
In part b) the right lead of dot 2 was  removed, i.e.,
parameters as in a) except for
$g_{2(12)}^{RR}=g^{LR}_{2(12)}=0$.
In part c) also the coupling of dot 1 is asymmetric:
  $g_{1}^{LL}=0.04,g_{1}^{RR}=0.06$, $g_{2}^{LL}=0.04$, $g_{2}^{RR}=0$, $g_{12}^{LL}=0.08$.\label{FigRGvertex1}}
\end{figure}

Let us now discuss the numerical solution
for the renormalized vertex functions of  Eq.\ (\ref{scaling2})/(\ref{scaling2level})
which is
presented in Figure\ \ref{FigRGvertex1}.
This  vertex function, a frequency-dependent renormalization of the
interactions $J,\tilde J$ between lead electrons and local states in the quantum dot,
 is  the fundamental theoretical 
quantity; it is not directly observable experimentally,
but -together with the occupations-
allows to calculate e.g.\ the current which will be discussed later.
In the following, all energies are normalized to the (equilibrium)
Kondo temperature for $\delta=0$, i.e., to  $T_K^0=T_K(\delta=0)$. \cite{footnoteTK0}
Due to the finite voltage $V$ applied, 
the two leads are thermalized by two different Fermi functions
with $\mu_L=\mu+V/2$ and $\mu_R=\mu-V/2$, respectively.
The quantum dot is out of equilibrium.

Here and in the following figures, the initial coupling constants 
which are not explicitly mentioned  follow from the ones of the Figure caption
since 
they stem from the same $t$'s of the Anderson impurity.
For example,  $g_{12}^{LR}=g_{12}^{LL}\sqrt{g_{1}^{RR}/g_{1}^{LL}}$.
We always set 
$\tilde{g}^{\lambda \lambda'}_i=0$ initially.

To understand the four peak structure of Figure\ \ref{FigRGvertex1}, it is instructive to
integrate Eq.\ (\ref{scaling2level})  analytically,
dividing the Eqs.\ by the squared coupling
constant from the left hand side 
(e.g.\ multiplying the first line by $1/g_i^2(\omega) $)
and afterwards keeping the $g$'s
on the right hand side
constant (at the initial values).
This is justified as long as the renormalization of the vertex functions stays small. But our  mere aim here is  to understand the positions and widths of the different peaks.
From Eq.\ (\ref{scaling2level}) we straightforwardly get in this way:
\begin{eqnarray}
g_{i}(\omega) &\sim& \frac{1}{\sum\limits_{\lambda}(\ln
\frac{|\omega+\lambda V/2|_{i}}{T^0_{k}}+\ln\frac{|\omega+i(\delta +\lambda V/2)|_{12}}{T^0_{k}})},\label{Eqdiscussion1} \\
g_{12}(\omega)&\sim&\frac{1}{\sum\limits_{i,\lambda}\ln\frac{|\omega+\frac{(\lambda V-i\delta)}{2}|_{12}}{T^0_{k}}},\label{Eqdiscussion2} \\
\tilde{g}_{i}(\omega)&\sim&\frac{1}{\sum\limits_{\lambda}\ln\frac{|\omega+i(\delta+\lambda
V/2)|_{12}}{T^0_{k}}},\label{Eqdiscussion3}
\end{eqnarray}
where $|\omega|_i\equiv\sqrt{\omega^2+\max^2\{\gamma_i,T\}}$, and prefactors of order
one (like $g_{12}/g_i$) before the logarithmic terms are neglected.

By means of Eq.\ (\ref{Eqdiscussion1}), we can now identify the origin
of the four Kondoesque peaks in the $g_i$'s seen in Figure\ \ref{FigRGvertex1}.
When the frequency matches the difference to the left or right 
Fermi level ($\omega=\pm V/2$) a single dot (spin) Kondoesque resonance develops;
 but
the flow to strong coupling is cut off by the corresponding decoherence
rate $\gamma_i$ (or $T$).
The other two peaks in $g_i$ stem from the orbital Kondo effect. Here, the localized state changes to the other orbital  and back to
the original orbital. These processes are enhanced
if the conduction electron scatters to the Fermi level of the
left or right lead, requiring  $\omega=-i \delta\pm V/2$.
Hence the resonances [and those of Eq.\ (\ref{Eqdiscussion3})]  are at $\omega=-i \delta\pm V/2$ and the cut-off is $\gamma_{12}$.
Similarly, the peaks for $g_{12}(\omega)$ in Eq.\ (\ref{Eqdiscussion2}) are at
 $\pm\frac{V\pm\delta}{2}$ (with $\delta/2$ instead of $\delta$ 
since $\omega$ is the average of incoming and outgoing frequency 
which  differ by $\delta$ as $g_{12}$ changes the orbital).

In part b) of  Figure\ \ref{FigRGvertex1} the right lead of dot 2 was
removed. As scattering to the removed lead is now prohibited,
one out of four peaks disappears for  $g_i$ and  $g_{12}$.
The missing peak is the one at the lowest value of $\omega$ in part a).
We can hence identify the missing peak with spin  ($g_2$)
and orbital scattering processes  ($g_1$ and  $g_{12}$) via  
the right lead of dot 2.
Part c)  of  Figure\ \ref{FigRGvertex1}
is for the situation where also dot 1 is coupled asymmetrically. Some peaks are clearly suppressed.

\begin{figure}[b]
{\includegraphics[width=7.5cm]{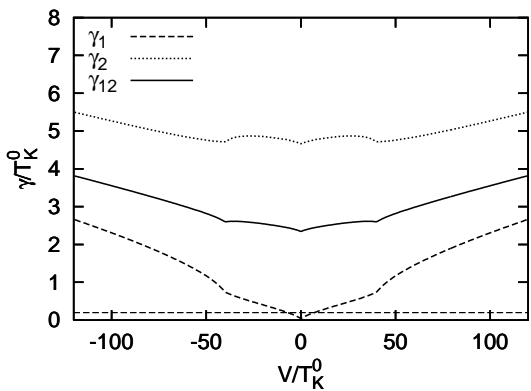}}
\caption{Scattering rates as a function of  bias voltage $V$ at $T/T^0_K=0.2$
  and $\delta/T^0_K=40$. The initial couplings are 
symmetric: $g_1^{\lambda \lambda'}=0.04$,
  $g_2^{\lambda \lambda'}=0.01$,
$g^{\lambda \lambda'}_{12} = 0.06$. For large bias voltage, the scattering rates are
  larger than $T$ (dashed line) but much smaller than $V$.
\label{FigGamma}}
\end{figure}

\subsection{Results: Decoherence rates}
\label{SecGamma}
Figure \ref{FigGamma}  shows the three decoherence rates $\gamma_1$, $\gamma_2$,
and $\gamma_{12}$ which need to be calculated self-consistently together
with the RG Eqs.
As in Eqs.\ (\ref{Eqdiscussion1})-(\ref{Eqdiscussion3}) it is
instructive to calculate  $\gamma$ from the unrenormalized $g$'s
(not self-consistently/in lowest order approximation).
At large orbital splitting $|\delta| \gg \max\{T,\frac{|V|}{4}\}$,
 this yields
for symmetric coupling to left/right lead
(Figure \ref{FigGamma} around  zero):
\begin{equation}
\begin{split}
\gamma_1 &\approx 4\pi g^2_1\max\{T,\frac{|V|}{4}\}, \\
\gamma_2 &\approx 4\pi g^2_{12}|\delta|, \\
\gamma_{12} &\approx 2\pi g^2_{12}|\delta|.
\end{split}
\end{equation}
The decoherence rate
for  dot 2, i.e., $\gamma_{2}$,  and the intra-orbital rate
$\gamma_{12}$ are both
proportional to $|\delta|$ as there are many decoherence processes,
reflecting the instability of the high energy level of dot 2.
Decoherence processes for dot 1 on the other hand become
only available at finite bias voltages.
 For very large  $|\delta|\gg T_K^0$,
the system shows the usual single-orbital (spin) Kondo effect.
Note that, for small $V$, we have $\gamma_1 <
T^0_K$. One might envisage
that this signals a flow to strong coupling and, hence,
a  break-down of the RG Eqs.
However, for the large splitting  $\delta$
the   real Kondo temperature is very much  reduced. It actually is only
$\approx 1/100 \; T^0_K$ here,  much smaller than the cut-off.

Let us now discuss the large voltage range of Figure \ref{FigGamma}.
For $|V| \gg |\delta|$, all curves show a similar asymptotic
behavior:
\begin{equation}
\begin{split}
\gamma_i &\approx 4\pi[(g_i)^2+(g_{12})^2]\max\{T,\frac{|V|}{4}\}, \\
\gamma_{12} &\approx
2\pi\sum\limits_i[\frac34(g_i)^2+(\tilde{g}_i)^2+(g_{12})^2]\max\{T,\frac{|V|}{4}\}.
\end{split}\label{EqGamma2}
\end{equation}
While all $\gamma_i$'s are now proportional to $|V|$,
we see in Figure \ref{FigGamma} that nonetheless $\gamma_i\ll |V|$.
This is due to the small prefactors $\sim g^2$ in Eq.\ (\ref{EqGamma2}).
The current which has
similar prefactors causes
the decoherence processes. 
For sufficiently large $V$, all  $\gamma_i$'s become
larger than temperature;
the current induced decoherence exceeds
the  temperature effect.

In between the two extremes $|V| \ll |\delta|$ and  $|V| \gg |\delta|$,
the transition from a strong
orbital decoherence rate of dot 2
 due to the high level energy to strong
current induced decoherences in both dots is clearly visible
at $V\approx\pm\delta$  in Figure \ref{FigGamma}.

Neglecting the finite orbital splitting, we can also get the
higher order asymptotic behavior for the decoherence
rates analogous to Ref.\ \onlinecite{Rosch01a}:
\begin{equation}
\gamma_i\sim
\frac{\pi}{4}\frac{V}{(\ln\frac{V}{T^0_K})^2}[1+\frac{2}{\ln\frac{V}{T^0_K}}+...].
\end{equation}

\subsection{Results: Non-equilibrium occupation}
\label{SecOccupation}

The numerical results for the different
 electron occupation of the two orbitals is 
shown in
Figure \ref{Fign}
for
different orbital splittings $\delta$. 
To ensure that the RG approach is valid, i.e., that we stay
in the weak coupling regime, we use here and in the following
a rather high temperature $T/T^0_K=2$.
Such temperatures are actually relevant in many experiments where often the lowest accessible temperatures are of the order of the
Kondo temperature.
At low bias voltage $V < \delta$, 
only the lower orbital is occupied for sufficient large
$\delta$, as is to be expected for $\delta
\gg T$. At high bias voltage $V>\delta$, the electrons tend
to stay in both dots with the same probability;  $\langle n_{1}-n_{2} \rangle$
decreases with increasing $V$.

\subsection{Results: Conductance for symmetric coupling}
\label{SecSym}

\begin{figure}[tb]
{\includegraphics[width=8.6cm]{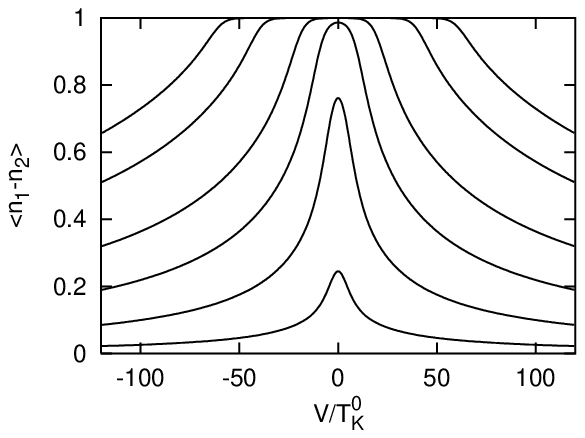}}
\vspace{-.8cm}

\caption{Difference of the occupation of the two orbitals,
$\langle n_1-n_2\rangle$, vs.\  bias voltage $V$ 
at different orbital splitting $\delta/T^0_K=60,40,20,10,4,1$
 (from top to bottom), temperature $T/T^0_K=2$, and symmetric 
bare couplings
  $g_i^{\lambda \lambda'}=0.04$,
$g^{\lambda \lambda'}_{12} = 0.06$.\label{Fign}}
\vspace{.4cm}

{\includegraphics[width=8.6cm]{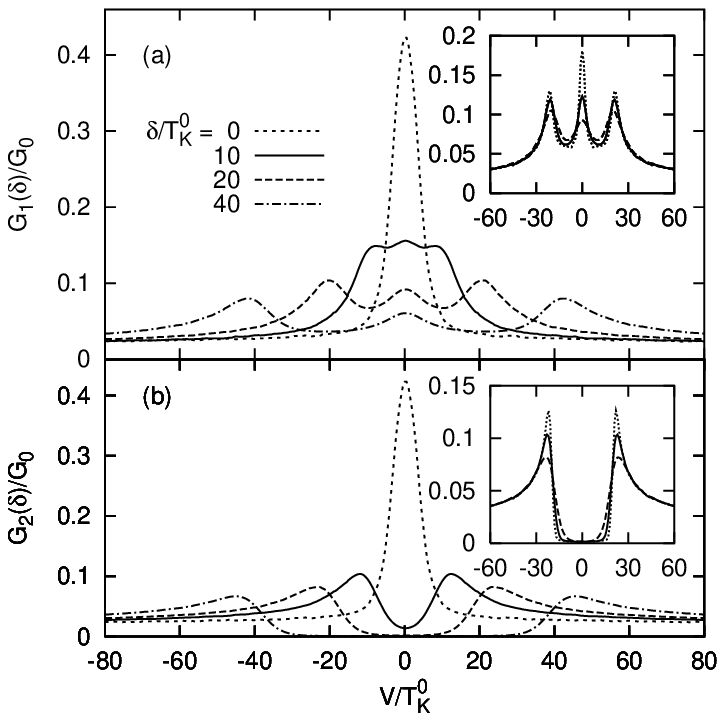}}
\vspace{-.6cm}

\caption{Differential conductance $G_i$ vs. bias voltage $V$ for symmetric
initial
  couplings $g_i^{\lambda \lambda'}=0.04$,
$g^{\lambda \lambda'}_{12} = 0.06$,
 temperature $T/T^0_K=2$, and different orbital splittings
  $\delta$.
Here and in the following figures, $G_0=2\frac{e^2}{h}$ denotes
the conductance quantum. 
Inset: temperature dependence of the differential conductance
for $\delta/T_K^0=20$ ($T/T^0_K=2$: dashed line; $T/T^0_K=1$: solid line ; $T/T^0_K=0.1$: dotted line).
\label{FigG1}}  
\end{figure}

Let us now turn to
the physical observable of interest, the  differential conductance through the two
dots
$G_i=\partial I_{i}(V)/ \partial V$ ($i=1,2$), presented in Figure \ref{FigG1}.
\begin{figure*}[tbh]
\begin{minipage}{12cm}
{\includegraphics[width=12.cm]{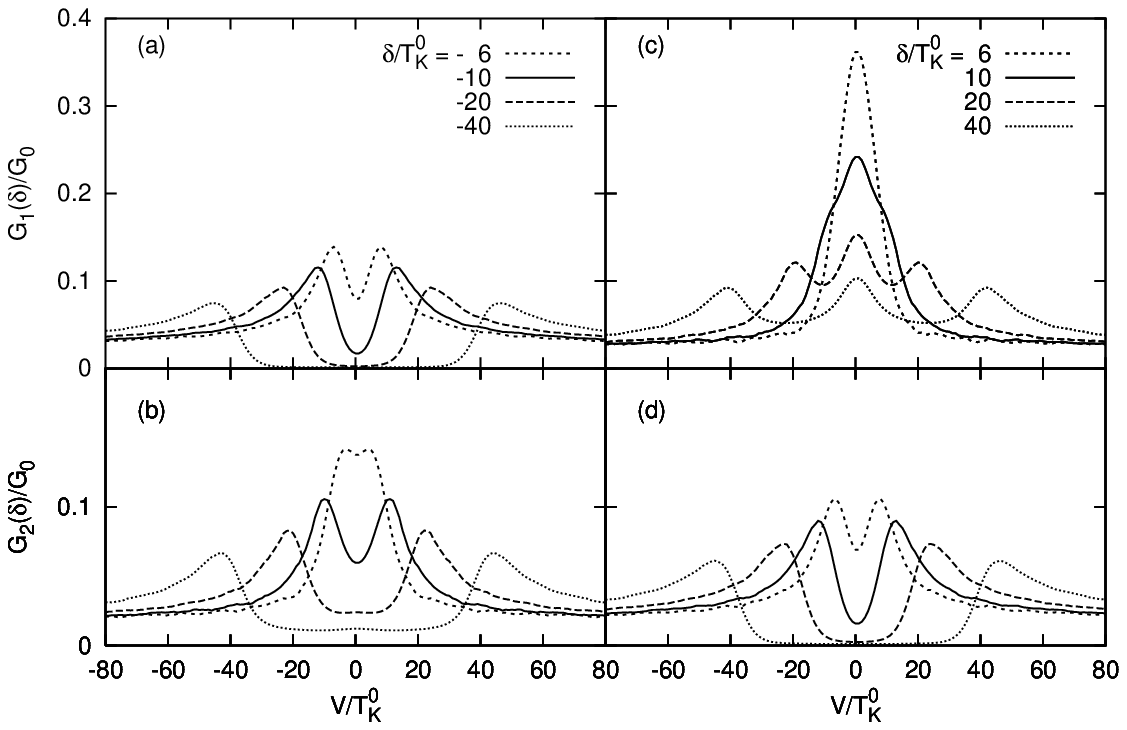}}
\end{minipage}\hfill
\begin{minipage}{4.3cm}
\caption{Differential conductance $G_i$ vs.\  bias voltage $V$ for symmetric
initial
couplings
  $g_1^{\lambda \lambda'}=0.04$,
  $g_2^{\lambda \lambda'}=0.01$,
$g^{\lambda \lambda'}_{12}=0.06$, temperature $T/T^0_K=2$, and
different orbital splittings
  $\delta$.
\label{FigG2}}  
\end{minipage}

\begin{minipage}{12cm}
{\includegraphics[width=12.cm]{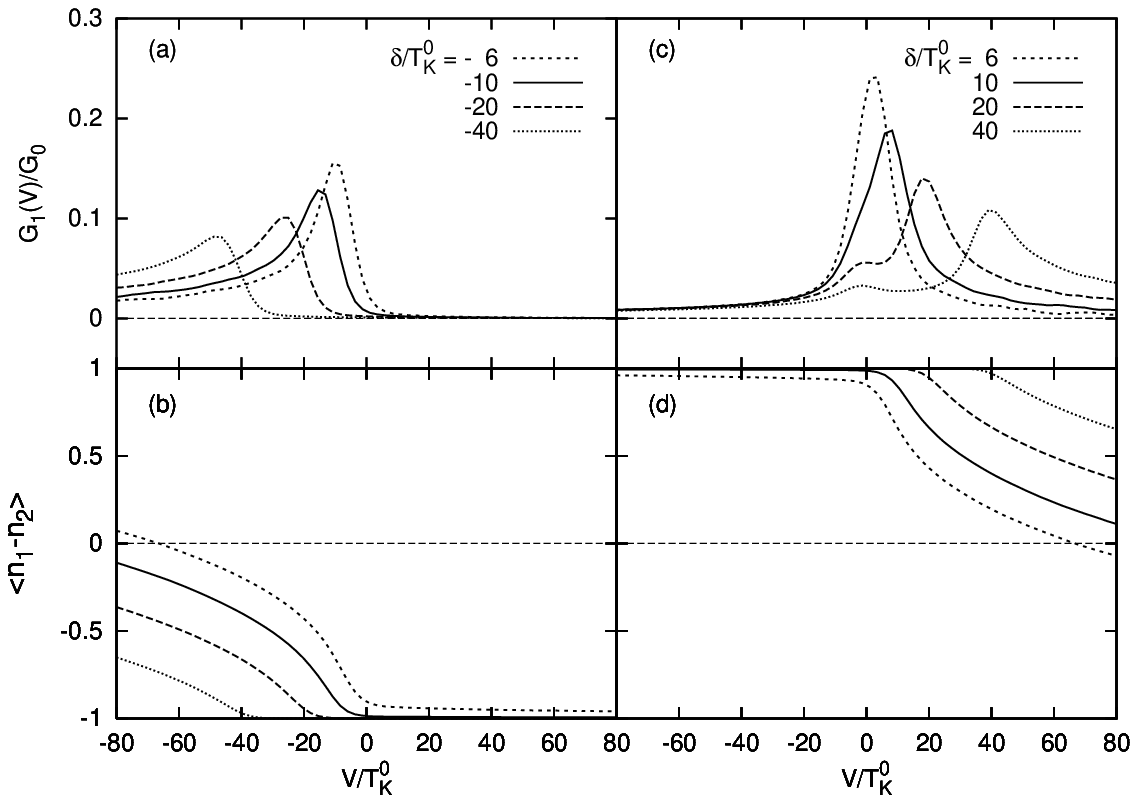}}
\end{minipage}\hfill
\begin{minipage}{4.3cm}
\caption{
Part (a) and (c)
are the same as in  Figure \ref{FigG2} but now for
asymmetric initial couplings 
 $g_1^{LL}=0.04$,  $g_1^{RR}=0.01$,
 $g_2^{LL}=0.01$,  $g_2^{RR}=0$, and  $g_{12}^{LL}=0.12$.
The Kondoesque peak at $V=-\delta$ is now missing.
Instead of the conductance $G_2$ which is zero,
part (b) and (d) show the
difference in the occupation of the two dots.
\label{FigG3} }
\end{minipage}
\end{figure*}
The conductance consists 
of the intra-dot spin scattering and the inter-dot
orbital scattering.
For $\delta=0$,
a pronounced peak is found at $V=0$,
characteristic of the orbitally enhanced Kondo effect. 
The flow to strong coupling is, however,  cut off
by temperature. If the orbital degeneracy
is lifted, the peak at $V=0$ is suppressed.
For  dot 2, it  finally disappears at large
$\delta$ when this dot is depopulated. For dot 1, we still have a spin  Kondo effect at large $\delta$, but with   a very much
reduced Kondo temperature $T_K^{\rm SU(2)} \ll T_K^0$.
Because also  $T_K^{\rm SU(2)} \ll T$, the peak at $V=0$  is very much smeared out.
At finite $\delta$, two new peaks occur at $V=\pm\delta$, corresponding
to orbital scattering.
For $T/T_K^0=2$, we study an interplay between 
temperature-induced decoherence and current-induced decoherence.
For small $\delta$ and $V$ the  decoherence is solely due to temperature,
whereas for larger the voltages it is mainly current noise.
For the $\delta=20$ and  $\delta=40$  curves in Figure \ref{FigG1},
the $\gamma_1$ decoherence rate is at least at the smaller voltages 
temperature-induced. In contrast, the important inter-orbital decoherence
rate $\gamma_{12}$ and also $\gamma_2$ are not even for smaller voltages.

If we decrease temperature, the main driving source for decoherence is the voltage-induced
current noise.
The change with temperature is studied in the inset of Figure \ref{FigG1},
showing the sharpening of the Kondoesque peaks.
At the lowest temperature $T/T_K^0=0.1$, we reached
the limit where the voltage-induced decoherence is dominating everywhere,
except for the central peak in  Figure \ref{FigG1}a).

 In Figure \ref{FigG2},
we show how the conductances change if the couplings
of the two dots to the leads are different.
In part c) and d), the stronger coupled dot (dot 1) is lower in energy.
The emerging
single-dot (spin) Kondo effect at $V\approx 0$ is somewhat suppressed,
see  Figure \ref{FigG2} c).
In contrast 
in part a) and b), it is the weaker coupled dot which 
is lower in energy (dot 2). Then
the spin Kondo peak  for dot 2 is
hardly discernible, see  Figure \ref{FigG2} d).

In the experiments of Ref.\ \onlinecite{Weisexp}, 
the inter-dot Coulomb interaction $U_{12}$ is much smaller than the intra-dot
Coulomb  interaction $U_{ii}$. Hence,
the initial inter-dot coupling is stronger
$g_{12} (\sim \frac{1}{U_{12}
+\epsilon_d}-\frac{1}{\epsilon_d})> g_i (\sim \frac{1}{U_{ii}
+\epsilon_d}-\frac{1}{\epsilon_d})$,
as in Figures \ref{FigG1} and \ref{FigG2}.

And, as in   Figures \ref{FigG1} and \ref{FigG2} b),
the single-dot (spin) Kondo peak at $V=0$ is much weaker
(if at all discernible) than the pronounced
orbital Kondoesque peaks  at $V\approx\pm\delta$.

\subsection{Results: Conductance for asymmetric coupling}
\label{SecAsym}
Let us now discuss the situation
of strongly different (initial/unrenormalized) couplings 
$g^{\lambda_1\lambda_2}_{m_1m_2}$ to the right and the left lead:
$g^{RR}_{m_1m_2}\neq g^{LL}_{m_1m_2}$.
This is typically the case in experiments,
\cite{Weisexp}
and can in extreme cases give rise to a very different
physical behavior compared to that of the symmetric coupling cases.

\begin{figure}[tb]
\begin{center}
{\includegraphics[width=8.cm]{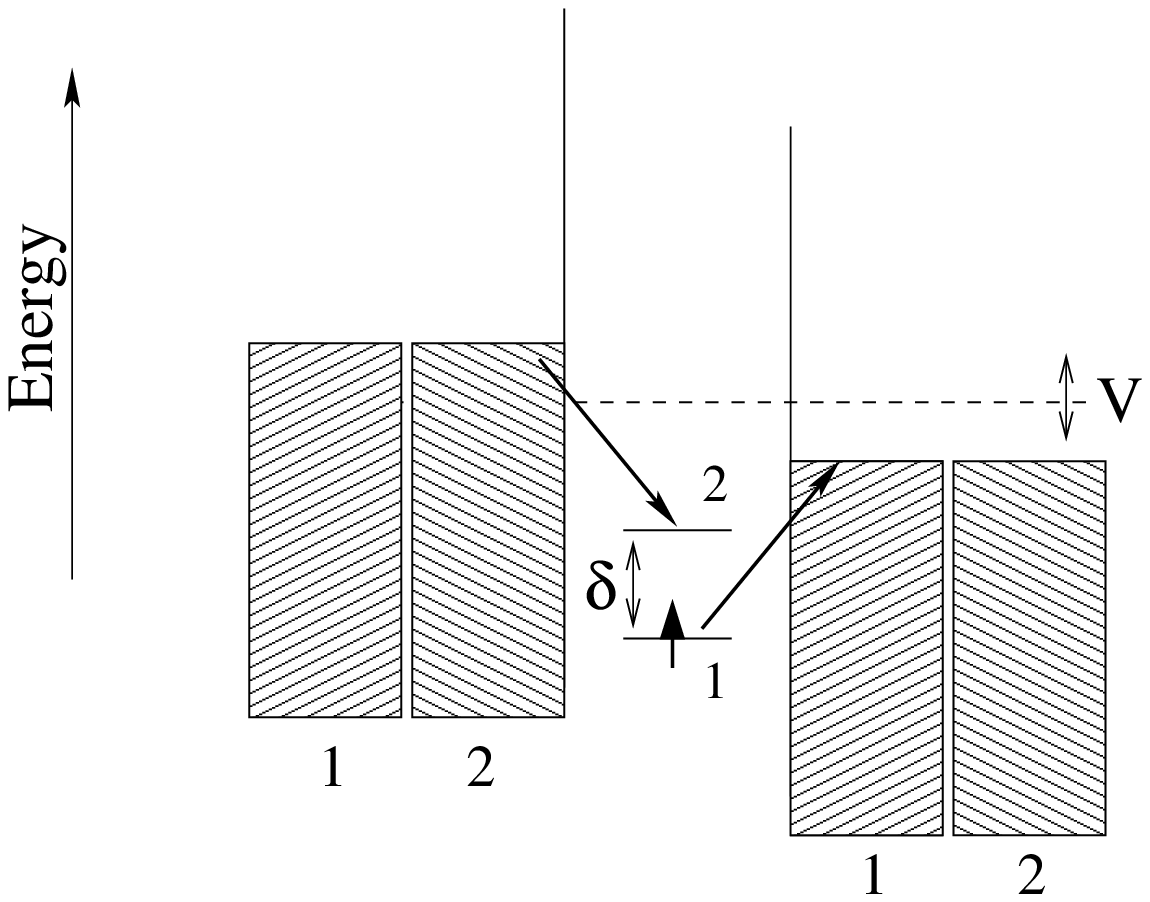}}
\end{center}
\caption{Scattering process which
connects the Fermi energy of
the left lead (at $+V/2$) and  dot 2 (upper level)
and that of the right lead  (at $-V/2$) with dot 1 (lower level);
there is a separate Fermi level for each dot as indicated.
 The  process displayed results, together with
the reverse scattering process,
 in a  Kondoesque resonance at $V\approx \delta$.
For $V>\delta$ the scattering process shown is
still possible but not the reverse one. 
\label{FigSketch}} 
\end{figure}

In Ref.\  \onlinecite{Weisexp}, this asymmetry is even
so strong that one of the dots is effectively decoupled
from the right lead  ($g^{RR}_{2(12)}=g^{LR}_{2(12)}=0$),
albeit still coupled to the left lead  via $g^{LL}_{2}$.
Such a situation is shown in Figure \ref{FigG3}.
Since dot 2 is decoupled from the left lead, the orbital Kondo
effect is now only possible at $V=+\delta$, not any longer at $V=-\delta$.
This can be understand by means of Figure \ref{FigSketch}.
At $V=+\delta$, we can move an electron from the Fermi energy of the
left lead into dot 2 and, conserving energy, the electron from dot
1 to the Fermi energy of the right lead, and vice versa. In the RG equations we hence get large couplings, resulting in an orbital Kondoesque peak.
If the coupling to the leads was symmetric we could
have similar scattering processes moving the electron from dot
2 to the right lead's Fermi energy and an electron from the
left lead's Fermi energy to dot 1 at $V=-\delta$.
However, since $g^{RR}_{2(12)}=0$ this second scattering
process is not possible, as was already argued in Ref.\ \onlinecite{Weisexp}.
Therefore, instead of two conductance peaks as in 
 Figures \ref{FigG1} and \ref{FigG2} we see only one at $V=+\delta$
in Figure \ref{FigG3}, which presents results for
 such strongly asymmetric couplings.
The nearly indiscernible spin Kondo peak and the missing
peak at $V\approx -\delta$ qualitatively agree with experiments.\cite{Weisexp}
As for symmetric coupling, the height of the orbital Kondo peak
at $V=\pm\delta$
decreases with increasing $\delta$.

\begin{figure}[tb]
{\includegraphics[width=8.6cm]{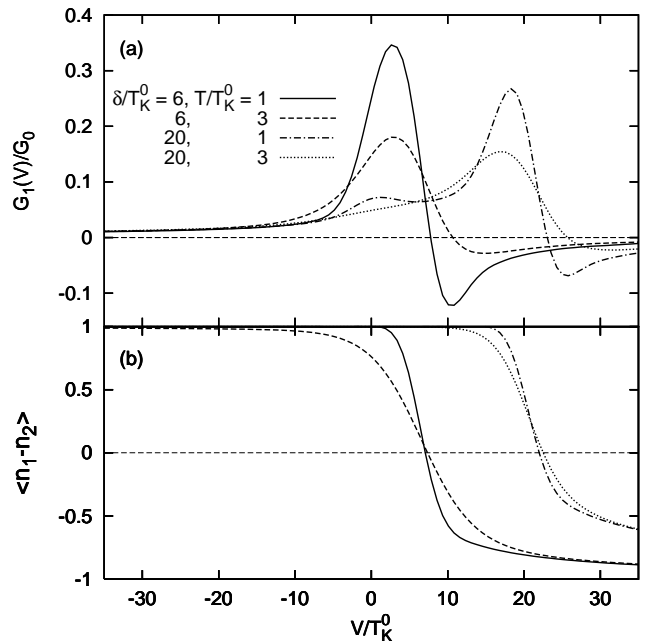}}
\caption{
Same as on the right hand side of Figure \ref{FigG3}
but now for
 $g_1^{LL}=0.01<g_1^{RR}=0.04$, and
$g_{12}^{LL}=0.06$.
($g_2^{LL}=0.01$,  $g_1^{RR}=0$ as before);
shown are two different values of $\delta$ and temperature.
For these parameters, the differential conductance
is negative after the Kondoesque peak at $V\approx\delta$
which can be explained by the inversion of the
occupation 
numbers, shown in part (b).
\label{FigG4} }
\end{figure}

An interesting phenomena in the case of asymmetric couplings
 occurs if  $g^{RR}_{2}=0$ as before but
$g^{RR}_1 \gg g^{LL}_1$:
As shown in Figure  \ref{FigG4}, we find a negative
differential conductance which is always  present after 
the orbital Kondo peak, i.e., for $V >\delta$.
Generally such negative conductances occur
when the hopping parameters
of the Anderson model fulfill
$|t_{R1}|\gg|t_{L1}|$, $|t_{R2}|\ll|t_{L2}|$.
How can we understand this unexpected 
 negative conductance?

The basic picture is that --although higher in energy--
dot 2 is increasingly occupied for  $V >\delta$.
Counterintuitively,
it becomes even more occupied than dot 1, see the lower part
of Figure  \ref{FigG4}.
The reason for this inversion of the occupation
(in comparison to the equilibrium occupation)
can be understood  from the sketch in
Figure \ref{FigSketch}.
The scattering process shown in the Figure is energetically
possible: Because $V-\delta>0$, 
we can transfer an electron from the Fermi level of the left
lead to the empty states above the Fermi level of the right lead
and still pay the energy $\delta$ to change
the local state  from dot 1
to dot 2. But
the reverse process, starting with an electron from the right lead's
Fermi level, is  not possible since $\delta-V<0$. All the other processes for
changing the local state back from dot 2 to dot 1 are much smaller because of 
$|t_{L1}|/|t_{R1}|\ll 1$ and $|t_{R2}|/|t_{L2}|\ll 1$.
Take for example the process involving the transfer of the dot 2 electron to
the right lead and simultaneously that of an electron from the left lead to dot
1. The amplitude for this process is by a factor of
$|g^{LR}_{12}/g^{LR}_{21}|\;(\approx |t_{L1}/t_{R1}||t_{R2}/t_{L2}|\ll 1)$
smaller than that of Figure 10. 
Hence, the localized state is nearly trapped in dot 2, $\langle
n_1-n_2\rangle<0$, for large bias voltage.

Altogether the behavior of the conductance in Figure  \ref{FigG4} can be 
explained as following:
At $V\approx \delta$ a new, inter-orbital channel opens for conductance
to the left lead of dot 1, the one displayed in Figure \ref{FigSketch}.
This explains the strong enhancement of $G_1$
for $V\approx \delta$; the orbital Kondo peak in 
Figure  \ref{FigG4}.
With a further increase of $V$, however, 
there is a dramatic decrease of $\langle n_1-n_2\rangle$ and hence
of the number of electrons in dot 1 ($\langle n_1\rangle$).
Because of this,
the contribution to the current
involving solely dot 1, i.e., the first line in Eq.\ (\ref{current}),
rapidly decreases. 
This decrease dominates for $V> \delta$,
resulting in a negative differential conductance.

Negative conductances for quantum dots with many levels have been
observed experimentally\cite{Johnson92} 
and described theoretically,\cite{Hettler02a}
 albeit in the sequential tunneling regime.
Our results show that similar effects are also possible
in the Kondo regime.
 
\subsection{Results: Linear conductance}
\label{SecLinear}

\begin{figure}[tb]
{\includegraphics[width=8.6cm]{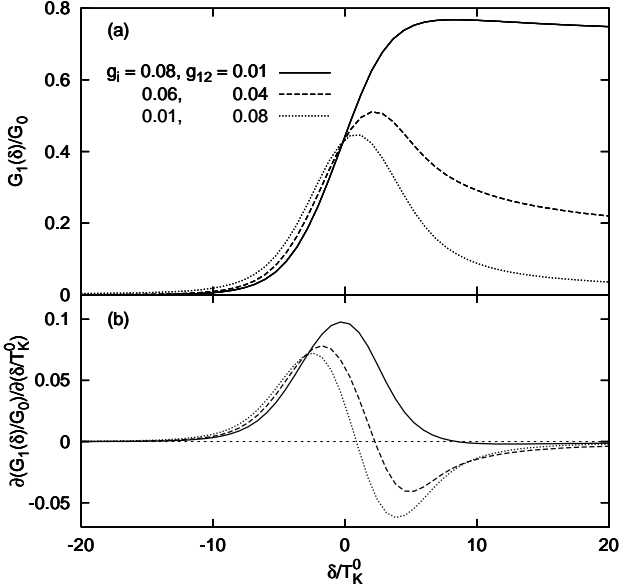}}
\caption{Part (a): Linear conductance $G_1(\delta)$ as a function of
 orbital splitting $\delta$ for symmetric coupling (see legend;
 $T/T_K^0=2$, note that $T_K^0$ depends on $g$).
Part (b)  shows $\partial G_1(\delta)/\partial \delta$
vs.\ $\delta$ 
which is symmetric/antisymmetric for a predominantly
spin/orbital Kondo effect. 
\label{FigLinear} }
\end{figure}

Substantial simplifications arise in the linear response regime for small
voltages since there are no nonequilibrium effects  any more like the
current induced decoherence rate $\gamma_i$ 
and the frequency dependence of the vertex.
This linear conductance can be used as an indicator
 whether we have a spin or an orbital Kondo effect,
corresponding to the limits  $g_i \gg g_{12}$
and  $g_{12} \gg g_i$, respectively.
In both cases, the linear conductance for dot 1
increases at $\delta \approx 0$, since then dot 1 becomes
occupied and conducting.
But while at larger values of $\delta$ ($\delta/T_K^0\gg 1$)
the spin Kondo effect still works, the orbital Kondo effect breaks down
in the linear conductance regime.
In Figure \ref{FigLinear} a), the conductance has hence a plateau form in the former case, while it has 
a peak form for the orbital Kondo effect.
The reason for this dramatic decrease of the conductance 
is simply that the RG flow of the inter-orbital coupling
(and that for the high energy dot) is cut-off at an energy
scale $\sim \delta$, preventing a strong renormalization and, hence,
suppressing the conductance at larger values of $\delta$.

This different behavior allows us to
identify the change of the conductance with $\delta$, i.e.,
$\partial{G(\delta)}/\partial{\delta}$,
as a good candidate to determine 
whether we have a spin or orbital Kondo effect.
As  Figure \ref{FigLinear} b) shows, this quantity is
symmetric for the spin Kondo effect but asymmetric 
for the orbital Kondo effect.
Note, that $\partial{G(\delta)}/\partial{\delta}$
 is often accessible experimentally;
the two dots need only be controllable by different gate voltages
for changing $\delta$.

\section{Conclusion}
\label{SecConclusion}
In conclusion, we reported on the extension of the nonequilibrium
perturbative RG scheme to  multi-orbital applications.
The method keeps the simplicity of the poor man's
scaling approach, but can be used for  nonequilibrium transport by  including
the frequency dependence of the vertex.
However, it is restricted to the weak coupling regime.
Either current or temperature has to cut off the flow to the strong coupling
fix point.

Specifically, we calculated  conductances
for the case of two quantum dots coupled 
only electrostatically. For SU(2) spin symmetry 
one spin (at $V\approx0$)  and two orbital Kondoesque peaks 
(at $V\approx \pm \delta$) are found, in agreement with
the expectations. For the typical experimental situation that
the intra-dot Coulomb interaction is much stronger than the inter-dot
Coulomb interaction (translating to initial couplings $g_{12}\gg g_1,g_2$)
however, the spin Kondo peak is much less pronounced,
if at all discernible.

For strongly asymmetric couplings with the extreme situation
that one of the dots is decoupled from one of the two leads,
the conductance peak at $V\approx-\delta$ disappears, as observed experimentally.\cite{Weisexp}
An unexpected result was the negative differential conductance
immediately after the remaining Kondoesque peak
at $V\approx \delta$,  occurring for initial couplings
 $g^{RR}_1\gg g^{LL}_1$ and
$g^{RR}_{2}\ll g^{LL}_{2}$.
This is a genuine nonequilibrium phenomenon.

Already in the the linear conductance regime, we can
 distinguish between spin and orbital Kondo effect
via the derivative 
$\partial{G_i(\delta)}/\partial{\delta}$ which
is symmetric and antisymmetric w.r.t. $\delta\rightarrow -\delta$,
respectively.

\section*{Acknowledgments}
We thank A.\ H\"ubel, A.\ A.\ Katanin, A.\ Rosch, D.\ Quirion, and J.\ Weis for discussions.
This work was supported in part by the Deutsche
Forschungsgemeinschaft through  the Emmy Noether program.

\bigskip

\appendix*
\section{DECOHERENCE RATES}

 In this Appendix,
we calculate the decoherence rates via 
the susceptibility or the corresponding correlation function:
\begin{equation}
\chi_{m_1m_2}(\omega)=\langle\!\langle X^{m_1m_2};X^{m_2m_1} \rangle\!\rangle_\omega,
\end{equation}
where $\langle\!\langle A;B  \rangle\!\rangle=-i \Theta(t) \langle
[A(t),B]\rangle$ and  $\langle\!\langle \ldots \rangle\!\rangle_{\omega}$
denotes the Fourier transformation to frequency space.
 
Near resonance $\omega_0$,  $\chi_{m_1m_2}(\omega)$
behaves like:
\begin{equation}
\chi_{m_1m_2}(\omega) \sim \frac{1}{(\omega-\omega_0) + i\gamma_{m_1m_2}},
\end{equation}
where $\gamma_{m_1m_2}$ is just the decoherence rate  wanted. 

For the exact diagrammatic evaluation at finite bias voltage, one should consider the
whole Keldysh contour
like in Ref.\ \onlinecite{Paaske04a}.
 However, for the lowest order contribution, it is
sufficient to  use only the equilibrium Green functions as in Ref.\ \onlinecite{Rosch03a}.
Moreover,  the proper decoherence rates to the lowest order
in $J$ can be obtained from the equation of motion method like in Ref.\ \onlinecite{Goetze71} 
with a truncation of the Eqs.\ in second
order in $J$.

The equation of motion of the correlation functions for Hamiltonian (\ref{Kondo}) is
written as:

\begin{eqnarray}
\lefteqn{\omega\chi_{m_1m_2}\!(\omega)\!=\!\langle[X^{m_1m_2}\!,\! X^{m_2 m_1}]\rangle\!+\!\langle\!\langle[X^{m_1m_2}\!,\!H];\!X^{m_2 m_1}\!\rangle\!\rangle_\omega}
\nonumber \\
&&=\!\langle X^{m_1m_1}\!\!-\!X^{m_2m_2}\rangle \!+\!(\epsilon_{m_2}\!\!-\!\epsilon_{m_1})\langle\!\langle X^{m_1m_2};\!X^{m_2 m_1}\rangle\!\rangle_\omega
\nonumber \\
&&+\!\!\! \!\!\sum\limits_{\lambda k;\lambda'm'k'}\!\! J^{\lambda\lambda'}_{m_2 m'}\langle\!\langle
X^{m_1m'}c^+_{\lambda' m' k'}c_{\lambda m_2k};X^{m_2 m_1}\rangle\!\rangle_\omega
\nonumber \\
&&-\!\!\!\!\!\sum\limits_{\lambda m k;\lambda'k'}\!\! J^{\lambda\lambda'}_{mm_1}\langle\!\langle
X^{mm_2}c^+_{\lambda'm_1k'}c_{\lambda mk};X^{m_2 m_1}\rangle\!\rangle_\omega
\nonumber \\
&&+\!\!\!\!\!\!\!\sum\limits_{\lambda k;\lambda'm'k'}\!\!\!\!\!\!(\tilde{J}^{\lambda\lambda'}_{m_2m'}\!-\!\tilde{J}^{\lambda\lambda'}_{m_1m'})\langle\!\langle
X^{m_1m_2}c^+_{\lambda'm'k'}c^{\phantom{+}}_{\lambda m'k};\!X^{m_2 m_1}\!\rangle\!\rangle_\omega.\nonumber
\end{eqnarray}
Here, we used the fact that $X^{m_1m_2}X^{m_3m_4}=\delta_{m_2m_3}X^{m_1m_4}$ in the
single electron subspace. The terms on the right hand side can  be
expressed in a similar way by another equation of motion. We then decouple 
this equation of motion  by using the approximation
\begin{eqnarray}
\lefteqn{\!\!\!\!\!\!\!\!\langle\!\langle X^{m_1m_2}c^+_1c_2c^+_3c_4;\!X^{m_2m_1}\rangle\!\rangle_\omega\approx} \nonumber\\&&\!\!\!\!\!\! [ \langle\! c_2c^+_3\rangle
\langle\! c^+_1c_4 \rangle \!+ \! \langle\! c_1^+c_2\rangle
\langle\! c^+_3c_4\rangle]
\langle\!\langle X^{m_1\!m_2}\!;\!X^{m_2\!m_1}\rangle\!\rangle_\omega.
\end{eqnarray}
This way, we neglect terms to order $J^3$, i.e.,
our decoherence rate is only valid to order $J^2$.
The   $\langle\! c_1^+c_2\rangle
\langle\! c^+_3c_4\rangle$ term can be absorbed in a renormalization of the
energies   
\begin{equation}
\epsilon_m\rightarrow\tilde{\epsilon}_m \approx \epsilon_m+\sum\limits_{\lambda k}
[J^{\lambda\lambda}_{mm}f_{\lambda}(\epsilon_{k})\!+\!\sum\limits_{m'}\tilde{J}^{\lambda\lambda}_{mm'}f_{\lambda}(\epsilon_{k})].
\end{equation}
We can effectively include this renormalization by denoting 
with  $\epsilon_m$ not the original level energy of the Hamiltonian
but the renormalized $\tilde{\epsilon}_m$ which is also
measured in the experiment. The RG Eqs.\ would include the same kind of change
if  higher order terms
were included. 

\begin{eqnarray}
\lefteqn{
\!\!\!\!\!\!\!\!\!\!\!\!\!\!\!(\omega\!+\!\epsilon_{m_1}\!-\!\epsilon_{m_2})\chi_{m_1m_2}\!(\omega)=\langle
X^{m_1m_1}\!\!-\!X^{m_2m_2}\!\rangle\!}\nonumber\\&&\;\;\;\;\;-i\pi B_{m_1m_2}\!(\omega)\chi_{m_1m_2}\!(\omega)\nonumber\\&&
\;\;\;\;\;+  i\pi\delta_{m_1m_2}\!\sum\limits_{m}C^{m}_{m_1}\!(\omega)
\langle\!\langle X^{mm};X^{m_1m_1} \!\rangle\!\rangle_\omega
\label{EqMotion}
\end{eqnarray}
plus an additional real part which typically results in a 
modification of $\tilde\epsilon$
but is of no interest for the decoherence rates since these
correspond to the imaginary part given by
\begin{widetext}
\begin{equation}
\begin{split} \label{epseps}
\!B_{m_1m_2}\!(\omega)=&\rho_0\!\!\!\!\sum\limits_{\lambda\lambda'km}\!\!\{J^{\lambda'\lambda}_{m_1m}J^{\lambda\lambda'}_{mm_1}f_{\lambda}(\epsilon_{k})[1\!-\!f_{\lambda'}(\epsilon_{k}\!-\!
\omega\!+\!\tilde{\epsilon}_{m_2}\!-\!\tilde{\epsilon}_{m})]
+[\tilde{J}^{\lambda'\lambda}_{m_1m}(\tilde{J}^{\lambda\lambda'}_{m_1m}-\tilde{J}^{\lambda\lambda'}_{m_2m})+
\delta_{mm_2}J^{\lambda'\lambda}_{m_1m_1}\\ &
(2\tilde{J}^{\lambda\lambda'}_{m_1m_1}\!-\!\tilde{J}^{\lambda\lambda'}_{mm_1})
\!-\!\delta_{mm_2}J^{\lambda'\lambda}_{m_2m_2}\tilde{J}^{\lambda\lambda'}_{m_1m}]f_{\lambda}(\epsilon_{k})[1\!-\!f_{\lambda'}(\epsilon_{k}-
\omega+\tilde{\epsilon}_{m_2}\!-\!\tilde{\epsilon}_{m_1}\!)]\}
+(m_1\!\leftrightarrow\! m_2,\omega \!\rightarrow\! -\omega),
\\
\!C^{m}_{m_1}(\omega)=&\rho_0\!\sum\limits_{\lambda\lambda'k}\!\!J^{\lambda'\lambda}_{mm_1}J^{\lambda\lambda'}_{m_1m}f_{\lambda}(\epsilon_{k})
(1-f_{\lambda'}(\epsilon_{k}+\omega+\tilde{\epsilon}_{m}-\tilde{\epsilon}_{m_1}))
+(\omega \rightarrow -\omega).
\end{split}
\end{equation}
\end{widetext}

For $m_1\neq m_2$, we directly 
obtain the decoherence rate 
from  Eq.\ (\ref{EqMotion}), taking $\omega\rightarrow\omega_0\approx \tilde{\epsilon}_{m_2}-\tilde{\epsilon}_{m_1}$:
\begin{equation}
\gamma_{m_1m_2}=\pi B_{m_1m_2}(\omega_0).
\label{EqGamma}
\end{equation}
This is the  rate entering the 
RG Eq.\ (\ref{scaling1}) as a cut-off.
In our perturbative RG calculation, we replace the bare couplings
$\rho_0J$  by the renormalized $g(\omega)$'s in Eq.\ (\ref{epseps}).
Then,
scaling corrections are included in a similar way as in the RG Eqs.\ themselves.

For $m_1=m_2$, the last term in Eq.\ (\ref{EqMotion})
cannot be neglected. 
The calculation becomes slightly more difficult since
a corresponding equation for
$\langle\!\langle X^{mm};X^{m_1m_1} \!\rangle\!\rangle_\omega$
(the terms entering the longitudinal susceptibility)
is needed.
However, often, one does not need the $m_1=m_2$
decoherence rate explicitly.
For example,  the calculations of the paper are for
a  SU(2) spin symmetry for both levels/dots.
Then, the longitudinal susceptibility and
decoherence rate 
$\gamma_{m_1m_1}$ with $m_1=\{i,\sigma\}$
is simply
equal to the transversal rate $\gamma_{m_1m_2}$
for  $m_2=\{i,-\sigma\}$.

As an example, the relaxation rate of the SU(N) model is
\begin{equation}
\gamma=2\pi N \rho_0^2 J^2 T,
\end{equation}
which can be obtained 
by simply substituting $J_{m_1m_2}=J$ and $\tilde{J}_{m_1m_2}=J/N$.

An additional check is the  spin-$\frac12$ Kondo model.
In a magnetic field and at a finite bias, we obtain from
Eq.\ (\ref{EqMotion}) and an analogous equation of motion for
$\langle\!\langle X^{mm};X^{m_1m_1} \!\rangle\!\rangle_\omega$
the transversal
and  longitudinal spin relaxation rates:
\begin{eqnarray}
\!\!\!\!\gamma_{\perp}\!&=\frac{\pi}{4} \rho_0^2J^2\!\sum\limits_{\lambda\lambda's}\!\int\!
d\epsilon&\!\!
\big[f_{\lambda}(\epsilon-\frac{s\delta}{2})(1-f_{\lambda'}(\epsilon+\frac{s\delta}{2}))
\nonumber\\&&
+f_{\lambda}(\epsilon)(1-f_{\lambda'}(\epsilon))\big],
\end{eqnarray}
\begin{eqnarray}
\!\!\!\!\!\!\!\!\!\!\!\!\!\!\gamma_{\parallel}\!&=\frac{\pi}{2} \rho_0^2J^2\!\sum\limits_{\lambda\lambda's}\!\int\!
d\epsilon &\!\!
\big[f_{\lambda}(\epsilon\!-\!\frac{s\delta}{2})(1\!-\!f_{\lambda'}(\epsilon\!+\!\frac{s\delta}{2}))
\big].
\end{eqnarray}
This is just the lowest order results one expects from 
nonequilibrium perturbation theory\cite{Paaske04a}, i.e.:
\begin{equation}
\gamma_{\perp} \approx\gamma_{\parallel}\approx4\pi \rho_0^2 J^2
\max\{T,\frac{V}{4}\}
\end{equation}
for $\max\{V,T\}\gg \delta$, and
\begin{equation}
\gamma_{\perp}\approx\frac{\gamma_{\parallel}}{2}\approx\pi
 \rho_0^2J^2\delta
\end{equation}
for $\max\{V,T\}\ll \delta$.


\begin{thebibliography}{99}
\bibitem{footnote1}
In second order scaling  theory [one-loop 
perturbative renormalization group (RG)]
the prefactor is given by the conduction electron bandwidth $D$
of the Coqblin-Schrieffer model.
It gets corrected by a factor $(N\rho_0J)^{(1/N)}$
in third order scaling theory 
or when calculating the Anderson impurity model 
 exactly by Bethe ansatz.\cite{Hewson}
\bibitem{NoteDiff}
A difference to Ref.\ \onlinecite{Rosch03a} is the way how we deal with the frequency conservation at the two vertices of Figure \ref{figFD}.
In  Ref.\ \onlinecite{Rosch03a}, the conduction electron
frequency of the intermediate state 
is set to the band edge $\omega'=\pm D$
so that the pseudo Fermion line (dashed line in Figure  \ref{figFD})
is off resonance, i.e., its frequency $\neq \epsilon_m$ where $m$ is the intermediate pseudo Fermion (localized) state.
In contrast in Eq.\ (\ref{scaling2}),  the localized state is
on resonance at $\epsilon_m$ so that the conduction electron
frequency $\omega'\neq D$. This is similar to the traditional poor man's scaling
approach where  $\omega'\neq D$ and the conduction electron momentum is such 
that $\epsilon_k=\pm D$, see, e.g.\ Ref.\ \onlinecite{Hewson} for details.
\bibitem{footnoteTK0} 
This is the Kondo temperature which 
the two dot system had if the orbital splitting was zero  but the $g$'s were the same. The reason for taking $T_K(\delta=0)$ as 
the energy unit is to study changes w.r.t.\ $\delta$,
keeping the other parameters constant.
\end{thebibliography}
\end{document}